\documentclass[english,prb,twocolumn]{revtex4-1}
\usepackage{ae,aecompl}
\usepackage[T1]{fontenc}
\usepackage[latin9]{inputenc}
\usepackage{babel}
\usepackage{float}
\usepackage{amsmath}
\usepackage{amssymb}
\usepackage{stmaryrd}
\usepackage{graphicx}
\usepackage[unicode=true,
 bookmarks=true,bookmarksnumbered=false,bookmarksopen=false,
 breaklinks=false,pdfborder={0 0 1},backref=false,colorlinks=false]
 {hyperref}

\makeatletter

\providecommand{\tabularnewline}{\\}

\makeatother

\begin{document}
\global\long\def\ket#1{\left|#1\right\rangle }%
\global\long\def\bra#1{\left\langle #1\right|}%
\global\long\def\braket#1#2{\langle#1|#2\rangle}%

\title{Quantum speed limit time in relativistic frame}
\author{N. A. Khan}
\email{niaz_phy@yahoo.com}

\affiliation{Centro de Física das Universidades do Minho e Porto~\\
Departamento de Física e Astronomia, Faculdade de Ciências, Universidade
do Porto, 4169-007 Porto, Portugal}
\author{Munsif Jan}
\email{mjansafi@mail.ustc.edu.cn}

\affiliation{Key Laboratory of Quantum Information of Chinese Academy of Sciences
(CAS), University of Science and Technology of China, Hefei 230026,
P. R. China }
\begin{abstract}
We investigate the roles of relativistic effect on the speed of evolution
of a quantum system coupled with amplitude damping channels. We find
that the relativistic effect speed-up the quantum evolution to a uniform
evolution speed of an open quantum systems for the damping parameter
$p_{\tau}\lesssim p_{\tau_{c0}}.$ Moreover, we point out a non-monotonic
behavior of the quantum speed limit time (QSLT) with acceleration
in the damping limit $p_{\tau_{c0}}\lesssim p_{\tau}\lesssim p_{\tau_{c1}},$
where the relativistic effect first speed-up and then slow down the
quantum evolution process of the damped system.  For the damping
strength $p_{\tau_{c1}}\lesssim p_{\tau}$, we observe a monotonic
increasing behavior of QSLT, leads to slow down the quantum evolution
of the damped system. In addition, we examine the roles of the relativistic
effect on the speed limit time for a system coupled with the phase
damping channels.
\end{abstract}
\maketitle

\section{Introduction}

The field of quantum information under relativistic constraints leads
to the emergence of a new field of high research intensity, known
as Relativistic Quantum Information (RQI). The most spectacular research
in this field have been devoted to the study of: entanglement between
quantum field modes in the accelerated frames \citep{Fuentes-Schuller2005,Alsing_2006,Bruschi2010,Wen_2014,SKSafi2014,Yasser2015,Qiang2018,Dong2018,Dong2019,Qiang2019,Dong_2019,Niaz2020,Shamsi2020,Sanchez_2020},
entanglement in black-hole space-times \citep{Emparan2006,Martin2010}\textbf{,}
entanglement of Dirac field in an expanding space-time \citep{Fuentes2010},
relativistic quantum metrology \citep{Ahmadi2014,Tian2015} and teleportation
with a uniformly accelerated partner \citep{Alsing2003}. The field
aims to understand the unification of the theory of relativity and
quantum information. One of the most fundamental manifestations of
RQI is the Unruh entanglement degradation between quantum field modes
in the accelerated frames \citep{Alsing_2006,Fuentes-Schuller2005,Yasser2015}
under single mode approximation. Moreover, the observer-dependent
property of entanglement has been successfully examined beyond the
single-mode approximation \citep{Bruschi2010}.

The tool of quantum information theory plays a prominent role in the
understanding of entanglement witness of polarization-entangled photon
pairs \citep{Fink2017}. It has been experimentally tested that the
photonic quantum entanglement persist in the accelerated frames. An
other promising experimental realization of relativistic scenario
is\textbf{ }``entanglement swapping protocol \citep{Nagele2020}'',\textbf{
} where a maximum Bell violation occurred in a suitable reference frame.

The minimal time required for the evolution of a quantum system is
known as ``Quantum Speed Limit Time'' (QSLT) \citep{Zhang2014}.
There exists a considerable amount of work dedicated to estimate the
minimal evolution time of a quantum system \citep{Zhang2014,Campaioli2019,Musadiq2019,Musadiq2020}.
For instance, QSLT was successfully investigated for the damped Jaynes-Cummings
and the Ohmic-like dephasing model \citep{Zhang2014}. Moreover, it
was found that the relativistic effect slow down the quantum evolution
of the qubit in the damped Jaynes-Cummings model. There have also
been studies involving the speed of quantum evolution of a single
free spin-$1/2$ particle coupled with phase damping channels in the
relativistic framework \citep{Niaz2015}. In addition, the nature
of QSLT in Schwarzschild space-time for the damped Jaynes-Cummings
and Ohmic-like dephasing models have been examined \citep{Haseli2019}.
Their results show that the QSLT decreased and increased by increasing
relative distance of quantum system to event horizon for damped Jaynes-Cummings
and Ohmic-like dephasing model, respectively.

Recently, the relativistic effects on the speed of quantum evolution
have been reported for a free Dirac field in non-inertial frames \citep{Xu2020}.
It is pointed out that the relativistic effects speed-up the evolution
of the quantum system coupled with the amplitude damping channels.
However, no relativistic effects have been encountered for the speed
of quantum evolution of the phase dapmed-system in a non-inertial
frame.

\begin{figure}[H]
\begin{centering}
\includegraphics[scale=0.5]{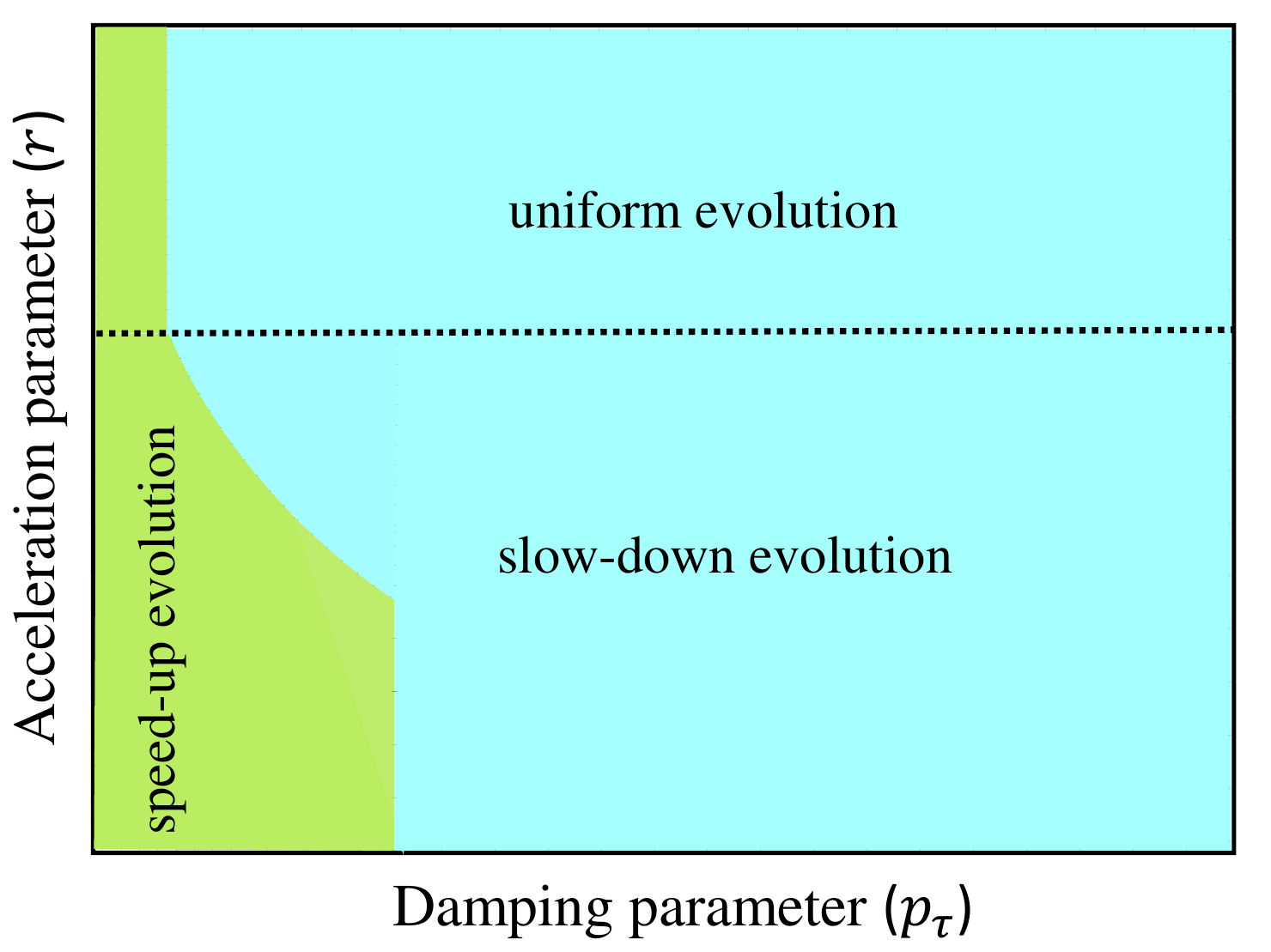}
\par\end{centering}
\caption{(Color online) The roles of relativistic effect on quantum speed limit
time of an open system coupled with the amplitude damping channels.\label{fig:RQSLTAD}}
\end{figure}

The aim of this article is to explore the role of relativistic effects
on the speed of evolution of a quantum system for a free scalar field
which manifests itself in the quantum noise. We point out that the
QSLT initially reduces to a minimum with increasing acceleration and
then trapped to a uniform fixed value for damping parameter $p_{\tau}\lesssim p_{\tau_{c0}}$.
This phenomenon leads to a speed-up of the quantum evolution initially,
and then reaches to a uniform evolution speed of an open quantum system.
In the region $p_{\tau_{c0}}\lesssim p_{\tau}\lesssim p_{\tau_{c1}}$,
 the QSLT first decreases to a minimum value, and then gradually
increases to a maximum uniform value as depicted in Fig.~\ref{fig:RQSLTAD}.

This
shows that the relativistic effect speed-up the quantum evolution
in the beginning and then slow down the speed of evolution of the
system. However, the quantum evolution of the system exhibits a slow
down behavior with increasing acceleration for $p_{\tau_{c1}}\lesssim p_{\tau}$,
leads to a larger QSLT in non-inertial frame. For each case, we notice
a uniform speed of evolution of the system in the large acceleration
limit, where the QSLT trapped to a fixed value. In addition, for the
phase damped-system, we obtain an acceleration independent speed limit
time in relativistic frame.

The structure of the paper is as follows. Sec.~\ref{sec:ScalarField},
is devoted to the theoretical background of the scalar field as observed
by uniformly accelerated observer. In particular, we review the mathematical
transformations between Minkowski and Rindler modes under single mode
approximation. In Sec.~\ref{sec:QSLT}, we present the physical scenario
and the mathematical procedure for calculating the QSLT of a quantum
system coupled with the amplitude damping channels in non-inertial
frames. Moreover, we analyze the relativistic effects on QSLT for
the amplitude damped open quantum system, when one observer move with
a uniform acceleration. In the last section, we sum up our conclusions.

\section{Scalar Field\label{sec:ScalarField}}

A real scalar field $\phi$ in two dimensional Minkowski space time
can be described by the massless Klein-Gordon equation, $\oblong\phi=0$.
This field can be expressed in terms of the positive and negative
energy solutions of the Klein-Gordon equation, given by \citep{Takagi1986,Bruschi2010}
\begin{equation}
\Phi_{\omega,\text{\ensuremath{\text{M}}}}=\int_{0}^{\infty}(a_{\omega,\text{\ensuremath{\text{M}}}}\varphi_{\omega,\text{M}}+a_{\omega,\text{\ensuremath{\text{M}}}}^{\dagger}\varphi_{\omega,\text{M}}^{*})d\omega
\end{equation}
where $a_{\omega,\text{\ensuremath{\text{M}}}}$ and $a_{\omega,\text{\ensuremath{\text{M}}}}^{\dagger}$
are the Minkowski annihilation and creation operators, obeying the
bosonic commutation relations. The positive-energy mode solution $\varphi_{\omega,\text{M}}$,
with respect to the timelike Killing vector field $\partial_{t}$,
for an inertial observer in Minkowski coordinates $(t,\,x),$ with
positive Minkowski frequency $\omega$ is given by
\begin{equation}
\varphi_{\omega,\text{M}}(t,x)=\frac{1}{\sqrt{4\pi\omega}}\exp\left[-i\omega(t-\varepsilon x)\right],\label{eq: energy-sol}
\end{equation}
where $\varepsilon$ can takes the value $1$ and $-1$ for modes
with positive (right movers) and negative (left movers) momentum,
respectively. The mode solutions satisfy the following relations
\begin{align}
(\varphi_{\omega,\text{M}},\varphi_{\omega^{'},\text{M}}) & =-(\varphi_{\omega,\text{M}}^{*},\varphi_{\omega^{'},\text{M}}^{*})=\delta_{\omega\omega^{'}},\nonumber \\
(\varphi_{\omega,\text{M}}^{*},\varphi_{\omega^{'},\text{M}}) & =0.\label{eq:commR}
\end{align}

The Klein-Gordon equation for a uniformly accelerated observer can
be more appropriately described by Rindler space-time. The Rindler
coordinates and Minkowski coordinates are related by \citep{Takagi1986,Bruschi2010}
\begin{equation}
\eta=\arctan\left(\frac{t}{x}\right),\qquad\chi=\sqrt{x^{2}-t^{2},}
\end{equation}
where $\chi=a^{-1}$, is the position and $\eta/a$ is the proper
time of the accelerated observer in region $\text{I}$. Here $a$
is a positive constant, referred as the acceleration of the uniformly
accelerated observer. The Rindler coordinates in region II can simply
be obtained by replacing $\eta=-\eta$. The Rindler coordinates have
ranges $0<\chi<\infty$ and $-\infty<\eta<\infty$.

The field can be expanded in terms of the energy solutions of the
Klein-Gordon equation in region $\text{I }$ and $\text{II}$ in the
Rindler coordinates is \citep{Takagi1986,Bruschi2010}
\begin{equation}
\Phi_{\Omega,R}=\int_{0}^{\infty}(a_{\Omega,\text{I}}\varphi_{\Omega,\text{I}}+a_{\Omega,\text{I}}^{\dagger}\varphi_{\Omega,\text{I}}^{*}+a_{\Omega,\text{II}}\varphi_{\Omega,\text{II}}+a_{\Omega,\text{II}}^{\dagger}\varphi_{\Omega,\text{II}}^{*})d\Omega,
\end{equation}
where $a_{\Omega,\sigma}$ and $a_{\Omega,\sigma}^{\dagger}$ are
the Rindler annihilation and creation operators for $\sigma\in\{\text{I, II}\}$,
respectively. It obeys the bosonic commutation relations. The $\varphi_{\Omega,\sigma}(t,x)$
is the positive frequency mode functions with respect to the timelike
Killing vector field $\pm\partial_{\eta}$, for the accelerated observer
in region $\sigma$, as given by
\begin{align}
\varphi_{\Omega,\text{I}}(t,x) & =\frac{1}{\sqrt{4\pi\Omega}}\left(\frac{x-\varepsilon t}{l_{\Omega}}\right)^{i\varepsilon\Omega},\nonumber \\
\varphi_{\Omega,\text{II}}(t,x) & =\frac{1}{\sqrt{4\pi\Omega}}\left(\frac{\varepsilon t-x}{l_{\Omega}}\right)^{-i\varepsilon\Omega},\label{eq:a}
\end{align}
where $\Omega$ is a dimensionless Rindler frequency, $l_{\Omega}$
is a positive constant of dimension length.

The field solution can also be expressed in the Unruh bases \citep{Takagi1986,Bruschi2010}
\begin{equation}
\Phi_{\Omega,\text{U}}=\int_{0}^{\infty}(a_{\Omega,\text{R}}\varphi_{\Omega,\text{R}}+a_{\Omega,\text{R}}^{\dagger}\varphi_{\Omega,\text{R}}^{*}+a_{\Omega,\text{L}}\varphi_{\Omega,\text{L}}+a_{\Omega,\text{L}}^{\dagger}\varphi_{\Omega,\text{L}}^{*})d\Omega,
\end{equation}
where $a_{\Omega,\nu}$ and $a_{\Omega,\nu}^{\dagger}$ are the Unruh
annihilation and creation operators for $\nu\in\{\text{R, L}\}$,
respectively, obey the bosonic commutation relations. The solution
of the field $\varphi_{\Omega,\nu}(t,x)$ in the Unruh bases are related
by
\begin{align}
\varphi_{\Omega,\text{R}} & =\cosh r_{\Omega}\varphi_{\Omega,\text{I}}+\sinh r_{\Omega}\varphi_{\Omega,\text{II}}^{*},\nonumber \\
\varphi_{\Omega,\text{L}} & =\cosh r_{\Omega}\varphi_{\Omega,\text{II}}+\sinh r_{\Omega}\varphi_{\Omega,\text{I}}^{*},\label{eq:}
\end{align}
The expression Eq.~\ref{eq:} shows the transformation between the
Unruh and the Rindler bases. Similarly, the transformation between
the Minkowski and the Unruh modes are related by \citep{Takagi1986,Bruschi2010}
\begin{equation}
\varphi_{\omega,\text{M}}(t,x)=\int_{0}^{\infty}\left(\alpha_{\omega\Omega}^{\text{R}}\varphi_{\Omega,\text{R}}+\alpha_{\omega\Omega}^{\text{L}}\varphi_{\Omega,\text{L}}\right)d\Omega,
\end{equation}
where
\begin{align}
\varphi_{\Omega,\nu}=\int_{0}^{\infty}\left(\alpha_{\omega\Omega}^{\nu}\right)^{*}\varphi_{\omega,\text{M}}d\omega,\quad\nu\in\{\text{R, L}\},\\
\alpha_{\omega\Omega}^{\text{R}}=\frac{1}{\sqrt{2\pi\omega}}\left(\omega l\right)^{i\varepsilon\Omega},\quad\alpha_{\omega\Omega}^{\text{L}}=\left(\alpha_{\omega\Omega}^{\text{R}}\right)^{*},
\end{align}
where $l$ is constant of dimension length.

A similar transformation between the Minkowski and the Unruh operators
may yield the following expression \citep{Takagi1986,Bruschi2010}
\begin{equation}
a_{\omega,\text{M}}=\int_{0}^{\infty}\left(\left(\alpha_{\omega\Omega}^{\text{R}}\right)^{*}a_{\Omega,\text{R}}+\left(\alpha_{\omega\Omega}^{\text{L}}\right)^{*}a_{\Omega,\text{L}}\right)d\Omega,
\end{equation}
where
\begin{equation}
a_{\Omega,\text{\ensuremath{\nu}}}=\int_{0}^{\infty}\alpha_{\omega\Omega}^{\nu}a_{\omega,\text{M}}d\omega,\quad\nu\in\{\text{R, L}\},
\end{equation}
One can obtain a relationship between the Minkowski and the Unruh
operators
\begin{align}
a_{\Omega,\text{\ensuremath{\text{I}}}} & =\cosh r_{\Omega}a_{\Omega,\text{\ensuremath{\text{R}}}}+\sinh r_{\Omega}a_{\Omega,\text{\ensuremath{\text{L}}}}^{\dagger},\nonumber \\
a_{\Omega,\text{\ensuremath{\text{II}}}} & =\cosh r_{\Omega}a_{\Omega,\text{\ensuremath{\text{L}}}}+\sinh r_{\Omega}a_{\Omega,\text{\ensuremath{\text{R}}}}^{\dagger}.\label{eq:-1}
\end{align}
Now we are in the position to relate the vacua and excited states
of the Minkowski, Rindler and Unruh modes. It is important to mention
that the Minkowski and the Unruh modes have common vacuum state, i.e.,
$\ket{0_{\Omega}}_{\text{M}}=\ket{0_{\Omega}}_{\text{U}}$. The Minkowski
vacuum state for a free scalar field can be expressed as a two-mode
squeezed state of the Rindler vacuum
\begin{equation}
\ket{0_{\Omega}}_{\text{M}}=\frac{1}{\cosh r}\sum_{n=0}^{\infty}\tanh^{n}r\ket{n_{\Omega}}_{\text{I}}\ket{n_{\Omega}}_{\text{II}},\label{eq:VacuumState}
\end{equation}
where $\ket{n_{\Omega}}_{\text{I}}$ and $\ket{n_{\Omega}}_{\text{II}}$
indicate the Rindler particle mode in region $\text{I}$ and $\text{II},$
respectively. Using single mode approximations, the Minkowski excited
state can be expressed as
\begin{equation}
\ket{1_{\Omega}}_{\text{M}}=\frac{1}{\cosh^{2}r}\sum_{n=0}^{\infty}\sqrt{n+1}\tanh^{n}r\ket{n+1_{\Omega}}_{\text{I}}\ket{n_{\Omega}}_{\text{II}},\label{eq:ExcitedStates}
\end{equation}
where $r$ is the acceleration parameter, defined as $\tanh r=\exp(-\pi\Omega)$,
with $\Omega=\left|k\right|c/a$, such that $0\leq r\leq\infty$ for
$0\leq a\leq\infty$. Here, $k$ is a wave vector denotes the modes
of the scalar field.

\section{Relativistic effects on the QSLT\label{sec:QSLT}}

In the following, we discuss the notion of a QSLT, which sets the
ultimate maximal speed of evolution of an open quantum system. It
basically determines the minimal time (lower bound) of evolution from
a mixed state $\varrho_{0}$ to its final mixed state $\varrho_{\tau}$.
A general expression for the QSLT of an open systems can be written
as \citep{Sun2019,Xu2020},
\begin{equation}
\tau_{QSL}=\frac{\left\Vert \varrho_{0}-\varrho_{\tau}\right\Vert _{\text{hs}}}{\overline{\left\Vert \dot{\varrho_{t}}\right\Vert }_{\text{hs}}},
\end{equation}
with
\begin{equation}
\overline{\left\Vert \dot{\varrho_{t}}\right\Vert }_{\text{hs}}=\frac{1}{\tau}\int_{0}^{\tau}dt\left\Vert \dot{\varrho_{t}}\right\Vert _{\text{hs}},
\end{equation}
where $\left\Vert \mathcal{O}\right\Vert _{\text{hs}}$ is the the
Hilbert-Schmidt norm of density operator $\mathcal{O}$. The $\left\Vert \mathcal{O}\right\Vert _{\text{hs}}=\sqrt{\sum_{i}e_{i}^{2}}$,
with $e_{i}$ being the singular values of $\mathcal{O}$. The term
in the numerator coined as the Euclidean distance $\mathcal{D}(\varrho_{0},\,\,\varrho_{\tau})=\left\Vert \varrho_{0}-\varrho_{\tau}\right\Vert _{\text{hs}}$.
It is important to mention that the QSLT turns out to be the actual
evolution time in the limit $\mathcal{D}(\varrho_{0},\,\,\varrho_{\tau})=\overline{\left\Vert \dot{\varrho_{t}}\right\Vert }_{\text{hs}}$.

We assume that the initial maximally entangled state between Alice
$\mathcal{A}$ and Bob $\mathcal{B}$ of single Minkowski mode $k$
reads;
\begin{align}
\ket{\psi} & =\frac{1}{\sqrt{2}}\left(\ket{0_{k}}_{\mathcal{A}}\ket{0_{k}}_{\mathcal{B}}+\ket{1_{k}}_{\mathcal{A}}\ket{1_{k}}_{\mathcal{B}}\right),\label{eq:BellState}
\end{align}
Given that the observer $\mathcal{B}$ undergoes a uniform acceleration
with respect to the inertial observer $\mathcal{A}$, the scalar particle
vacuum and excited state of $\mathcal{B}$ in Minkowski space are
transformed into two causally disconnected Rindler regions $\text{\ensuremath{\text{I}}}$
and $\text{II}$ for particles and anti-particles, respectively. Using
single mode approximation, the state (Eq.~\ref{eq:BellState}) can
be expressed in terms of Rindler states of the free scalar field,
as given by
\begin{align}
\ket{\Psi} & =\frac{1}{\sqrt{2}\cosh r}{\displaystyle \sum_{n=0}^{\infty}}\tanh^{n}r(\left|0nn\right\rangle +\frac{{\displaystyle \sqrt{n+1}}}{\cosh r}\left|1n+1n\right\rangle ),\label{eq:BState-Acc-SF-1}
\end{align}
where $\left|xyz\right\rangle =\ket{x_{k}}_{\mathcal{A}}\ket{y_{k}}_{\mathcal{B_{\text{I}}}}\ket{z_{-k}}_{\mathcal{B_{\text{II}}}}$
for $x\in(0,\,1)$ and $y,z\in(n,\,n+1)$. 

In what follows, we study the evolution of the amplitude-damped quantum
system in the relativistic framework (especially Rindler basis in
region I). Taking the trace over the inaccessible state of region
II, we obtain a reduced density operator , $\rho_{\mathcal{AB}_{\text{I}}}$,
of Alice and physically accessible region of Bob, as given by
\begin{align}
\rho_{\mathcal{AB}_{\text{I}}} & =\ket 0\bra 0\otimes\mathcal{M}_{nn}\,\,+\,\,\left|1\right\rangle \left\langle 1\right|\otimes\mathcal{M}_{n+1n+1}\nonumber \\
 & \,\,+\left|1\right\rangle \left\langle 0\right|\otimes\mathcal{M}_{n+1n}+\left|0\right\rangle \left\langle 1\right|\otimes\mathcal{M}_{nn+1},\label{eq:ReducedMatrix-SF}
\end{align}
where
\begin{align}
\mathcal{M}_{nn} & =\frac{1}{2\cosh^{2}r}{\displaystyle \sum_{n=0}^{\infty}}\tanh^{2n}r\ket n\bra n,\nonumber \\
\mathcal{M}_{n+1n} & =\frac{1}{2\cosh^{3}r}{\displaystyle \sum_{n=0}^{\infty}}\sqrt{n+1}\tanh^{2n}r\ket{n+1}\bra n,\nonumber \\
\mathcal{M}_{n+1n+1} & =\frac{1}{2\cosh^{4}r}{\displaystyle \sum_{n=0}^{\infty}}(n+1)\tanh^{2n}r\ket{n+1}\bra{n+1},\nonumber \\
\mathcal{M}_{nn+1} & =\mathcal{M^{*}}_{n+1n}.\label{eq:MMMM}
\end{align}
Let consider the inertial observer $\mathcal{A}$ of the system is
under the action of an amplitude damping channel. In this case, nothing
happens, if the system is in the ground state. However, it change
the dynamic of the excited state of the system. The Krauss operators
$\mathcal{K}_{i}$ for $i\in(0,1)$ of this model is represented by
the following positive quantum map \citep{Nielsen2000}
\begin{equation}
\mathcal{K}_{0}=\left[\begin{array}{cc}
1 & 0\\
0 & \sqrt{p_{t}}
\end{array}\right],\quad\mathcal{K}_{1}=\left[\begin{array}{cc}
0 & \sqrt{1-p_{t}}\\
0 & 0
\end{array}\right],
\end{equation}
where $p_{t}=\exp\left(-\Gamma t\right),$ is a damping parameter,
an exponential decay of the excited population with decay rate $\Gamma$.
The system-environment interaction can be represented by the Krauss
decomposition of the quantum channel, $\mathcal{E}(\rho_{0})=\sum_{i}\mathcal{K}_{i}^{\dagger}\rho_{0}\mathcal{K}_{i}$,
satisfying $\sum_{i}\mathcal{K}_{i}^{\dagger}\mathcal{K}_{i}=\text{I}$.
Thus, the reduced density operator Eq.~\ref{eq:ReducedMatrix-SF},
after inertial observer interacting with amplitude damping channel
is given by
\begin{align}
\rho_{\mathcal{AB}_{\text{I}}}(p_{t}) & =\ket 0\bra 0\otimes\mathcal{M}_{nn}+\sqrt{p_{t}}(\left|1\right\rangle \left\langle 0\right|\otimes\mathcal{M}_{n+1n}+\text{h.c})\nonumber \\
 & \,\,\,+(p_{t}\left|1\right\rangle \left\langle 1\right|+(1-p_{t})\left|0\right\rangle \left\langle 0\right|)\otimes\mathcal{M}_{n+1n+1},\label{eq:ReducedMatrix-AD-Sf}
\end{align}
In order to characterize the speed of evolution for the amplitude
damped quantum systems, we first calculate the Euclidean distance
$\mathcal{D}(p_{\tau},\,\,r),$ between the noise-free initial state
$\rho_{\mathcal{AB}_{\text{I}}}$ to its amplitude decoherence state
$\rho_{\mathcal{AB}_{\text{I}}}(p_{\tau})$ in the accelerated frame.
Calculating the singular values of the reduced system $\rho_{\mathcal{AB}_{\text{I}}}-\rho_{\mathcal{AB}_{\text{I}}}(p_{\tau})$
and taking the square root of the sum of square of them, one finds
that the Euclidean distance,
\begin{equation}
\mathcal{D}(p_{\tau},\,\,r)=\frac{1-\sqrt{p_{\tau}}}{\sqrt{2}}\sqrt{\left(1+\sqrt{p_{\tau}}\right)^{2}+a^{2}(r)},\label{eq:EDistance}
\end{equation}
where $a(r)$ is the acceleration dependent parameter, given by
\begin{align}
a(r) & =\frac{1}{\cosh^{3}r}{\displaystyle \sum_{n=0}^{\infty}}\sqrt{n+1}\tanh^{2n}r,\nonumber \\
 & =\frac{1}{\cosh r\sinh^{2}r}\text{Li}_{-\frac{1}{2}}\left(\tanh^{2}r\right).\label{eq:Constant}
\end{align}
The trace norm of the $\dot{\rho}_{\mathcal{AB}_{\text{I}}}(p_{\tau})$
can be obtained as
\begin{equation}
\overline{\left\Vert \dot{\rho}_{\mathcal{AB}_{\text{I}}}(p_{\tau})\right\Vert }_{\text{hs}}=-\frac{1}{2\sqrt{2}\tau}\int_{1}^{p_{\tau}}dp_{t}\sqrt{\frac{4p_{t}+a^{2}(r)}{p_{t}}},\label{eq:RhoDot0}
\end{equation}
\begin{figure}[H]
\begin{centering}
\includegraphics[scale=0.34]{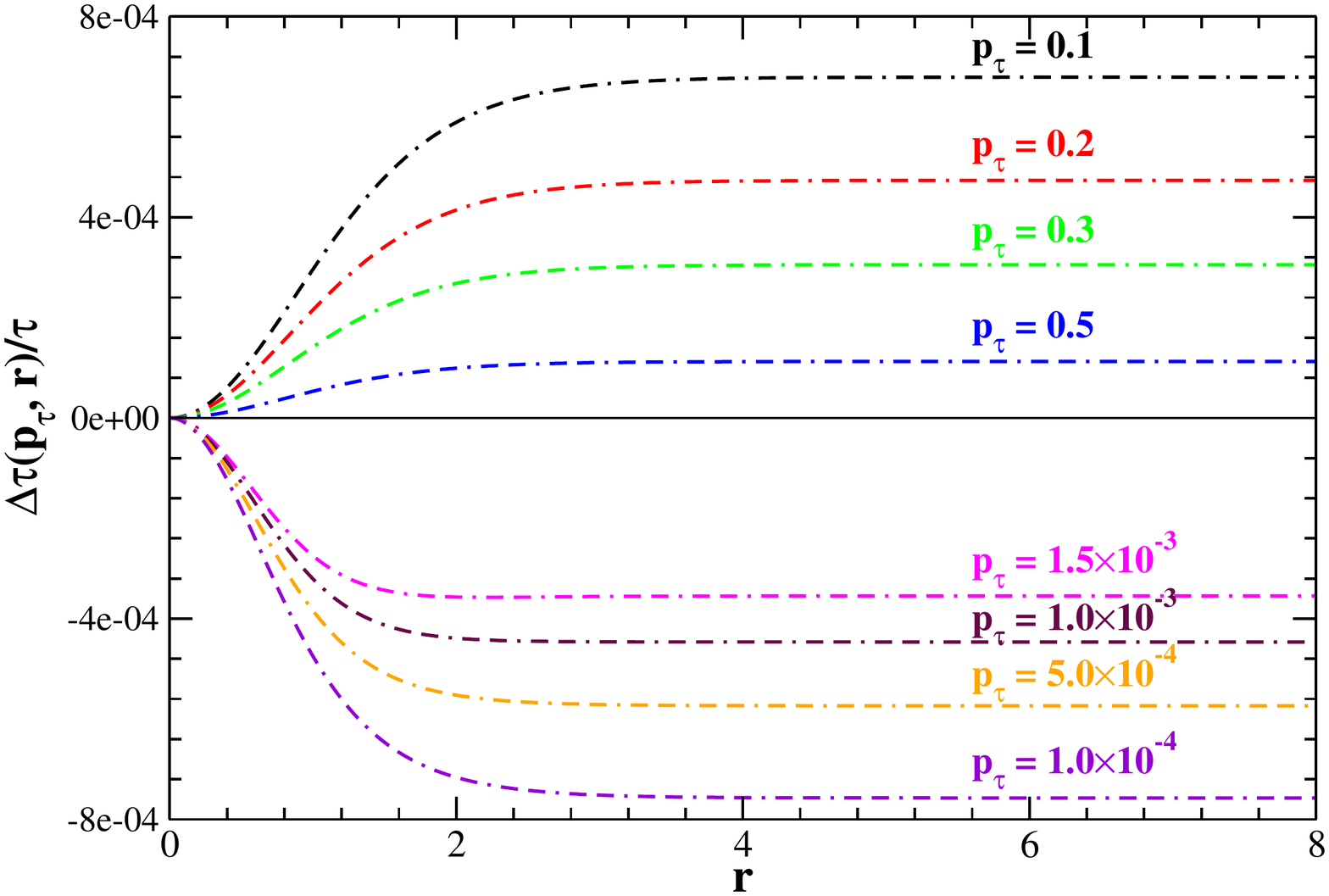}
\par\end{centering}
\caption{(Color online) The $\Delta\tau(p_{\tau},\,\,r)$ as a function of
acceleration parameter $r$ of the accelerated bosonic observer (scalar
field) for different values of damping strength $p_{\tau}$ for $p_{\tau}\protect\geq0.1\,\,\&\,\,p_{\tau}\protect\leq1.5\times10^{-3}$
of the amplitude damping channels.\label{fig:ScalerF}}
\end{figure}
\begin{figure}[H]
\begin{centering}
\includegraphics[scale=0.34]{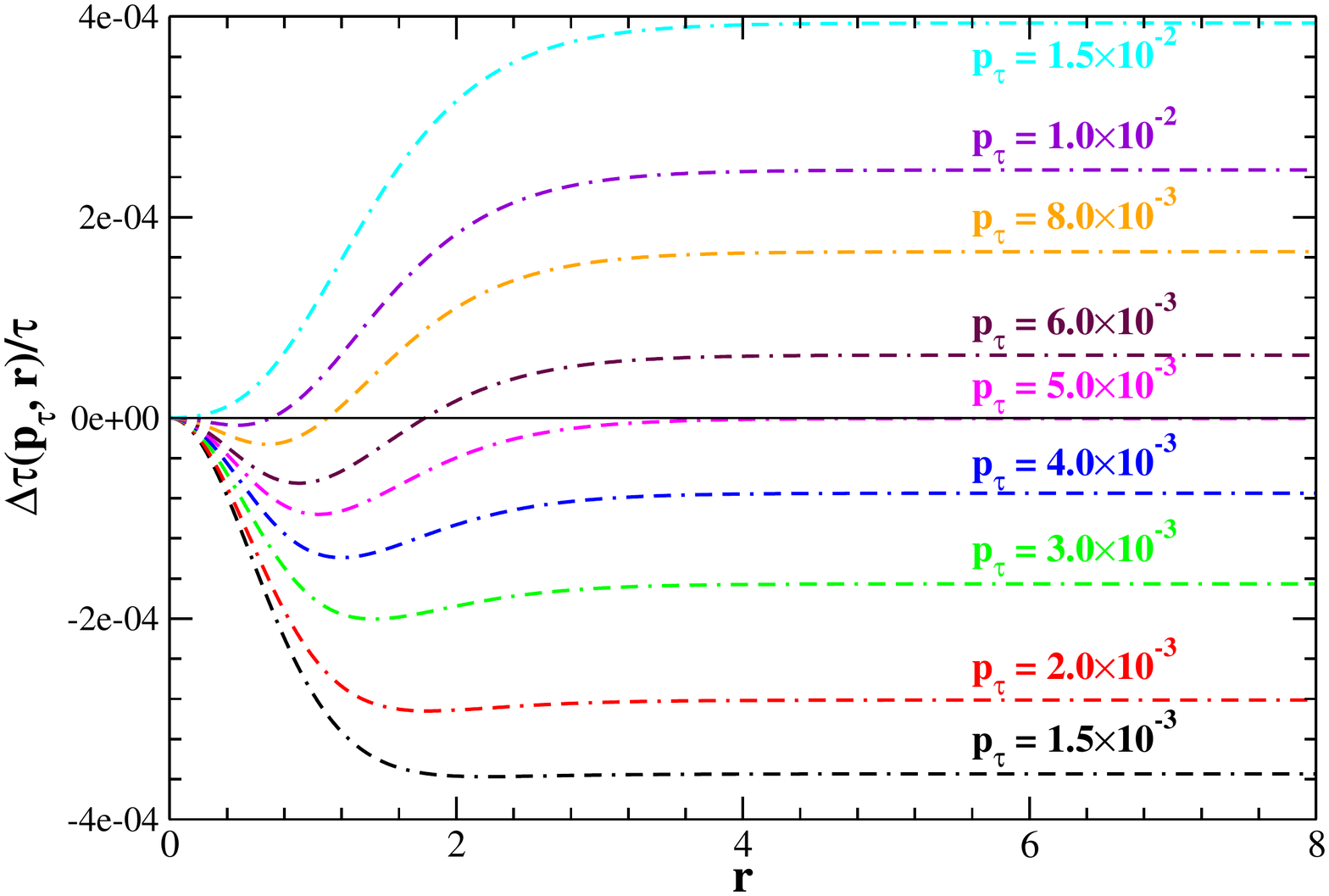}
\par\end{centering}
\caption{(Color online) The $\Delta\tau(r,\,\,p_{\tau})$ as a function of
acceleration parameter $r$ of the accelerated bosonic observer (scalar
field) for different values of damping strength $p_{\tau}$ for $1.5\times10^{-2}\protect\geq p_{\tau}\protect\geq1.5\times10^{-3}$
of the amplitude damping channels.\label{fig:ScalerF-2}}
\end{figure}
The expression for quantum speed limit time \textbf{$\tau(\,p_{\tau},\,\,r)$},
of the amplitude-damped state in the relativistic frame has the form
\begin{equation}
\tau(p_{\tau},\,r)=\frac{2\tau(1-\sqrt{p_{\tau}})\sqrt{\left(1+\sqrt{p_{\tau}}\right)^{2}+a^{2}(r)}}{-\int_{1}^{p_{\tau}}dp_{t}\sqrt{\frac{4p_{t}+a^{2}(r)}{p_{t}}}},\label{eq:RQSLT}
\end{equation}
The integral in the denominator is given by\footnote{The integral has the simplified form
\begin{align*}
-\int_{1}^{p_{\tau}}dp_{t}\sqrt{\frac{4p_{t}+a^{2}(r)}{p_{t}}} & =\sqrt{4+a^{2}(r)}-\sqrt{p_{\tau}\left(4p_{\tau}+a^{2}(r)\right)}\\
 & -\frac{1}{2}a^{2}(r)(\sinh^{-1}\frac{2\sqrt{p_{\tau}}}{a(r)}-\sinh^{-1}\frac{2}{a(r)}).
\end{align*}
}. It is straightforward to show that the QSLT $\tau(p_{\tau},\,r),$
reduce to pure Unruh decoherence form of the scalar field in the absence
of quantum noise $(p_{\tau}=0)$. In this limit, the $\tau(p_{\tau}=0,\,r),$
only depends on the acceleration parameter of the accelerated observer.
On the other hand, the $\tau(p_{\tau},\,r=0)$, represents the QSLT
of the purely amplitude damped quantum system in the non-relativistic
frame.

Our analysis of the quantum system coupled with the amplitude damping
channel indicates that the relativistic effect may speed-up or slow
down the quantum evolution of the damped-system, leads to a smaller
or larger $\tau(p_{\tau},\,r)$, respectively, in non-inertial frame
as illustrated in Fig.~\ref{fig:ScalerF}. It is shown that the QSLT
$\Delta\tau(p_{\tau},\,r)=\tau(p_{\tau},\,r)-\tau(p_{\tau},\,r=0)$,
is monotonically an increasing and a decreasing function of acceleration
parameter $r$ for $p_{\tau}\geq0.1$ and $p_{\tau}\leq p_{\tau_{c0}}$,
respectively. Hence, the relativistic effect slow down the speed of
evolution in the former case, whereas in the later case speed-up the
evolution of the quantum system. In the limit $p_{\tau}\sim1,$ there
exist no relativistic effects on the speed of evolution of strongly
damped system i.e., $\tau(p_{\tau},\,r)\approx\tau(p_{\tau},\,r=0)$,
where the $\Delta\tau(p_{\tau},\,r)$ approaches to a vanishingly
small value. On the other hand, the relativistic effect speed-up the
quantum evolution of the weakly damped system ($p_{\tau}\leq p_{\tau_{c1}}$),
results a smaller $\Delta\tau(p_{\tau},\,r)$ in the non-inertial
frame. It is important to mention that the $\Delta\tau(p_{\tau},\,r)$
for various $p_{\tau}$ gradually approach to a fixed value, leads
to a uniform evolution speed of the system.
\begin{table}
\begin{centering}
\begin{tabular}{|c|c|}
\hline 
Critical damping parameters & Numerical values\tabularnewline
\hline 
\hline 
$p_{\tau_{c0}}$ & $1.5\times10^{-3}$\tabularnewline
\hline 
$p_{\tau_{c1}}$ & $1.5\times10^{-2}$\tabularnewline
\hline 
\end{tabular}
\par\end{centering}
\caption{Numerical values of the critical damping parameter, where the QSLT
shows different behavior with respect to acceleration.\label{tab:table}}
\end{table}

The anomalous behavior of QSLT under Unruh decoherence as a function
of acceleration parameter for various damping strength, on the interval
$p_{\tau}\in(p_{\tau_{c1}},p_{\tau_{c0}})$ are depicted in Fig.~\ref{fig:ScalerF-2}.
One can clearly see that the QSLT decreases to a minimum value in
the beginning, then increases to a fixed value with increasing acceleration.
This shows that the relativistic effect first speed-up the evolution
process, then exhibit a gradual deceleration process to a uniform
evolution of the system. Furthermore, we notice an acceleration independent
behavior of QSLT for $p_{\tau}\approx5\times10^{-3},$ in the large
acceleration limit.

The anomalous ---monotonic and non-monotonic--- behavior of QSLT
is due to the competition between accelerated parameter $r$ and damping
parameter $p_{\tau}$. It is easy to see an enhancement of the QSLT
with acceleration for $p_{\tau}\geq0.1$, trapped to a fixed value
for the large acceleration limit. However, this behavior ceased out
for the highly damped limit $p_{\tau}\sim1$, where the QSLT turns
out to be acceleration independent. In this limit, we may say that
the system is completely evolved. For a weakly coupled system $(p_{\tau}\ll5\times10^{-3})$,
the acceleration parameter dominates, leads to the degradation of
QSLT in non-inertial frames. In particular, the QSLT of the system
in the absence of noise turns out:
\begin{equation}
\tau(p_{\tau}=0,\,r)=\frac{2\sqrt{a^{2}(r)+1}}{\sqrt{a^{2}(r)+4}+\frac{1}{2}a^{2}(r)\sinh^{-1}\frac{2}{a(r)}}.
\end{equation}
In addition, on the interval $p_{\tau}\in(p_{\tau_{c1}},p_{\tau_{c0}})$,
first the acceleration dominates the damping parameter, then damping
parameter dominates the acceleration, results in decreasing and increasing
of QSLT of the open quantum system.

Furthermore, the influence of the relativistic effects on the QSLT
for the quantum systems coupled with the phase damping channels in
non-inertial frames has also been examined. The relativistic QSLT
can be calculated as
\begin{equation}
\tau(q_{\tau})=\frac{1-\sqrt{1-q_{\tau}}}{1-\sqrt{q_{\tau}}}.\label{eq:QSLT-PhaseDaming}
\end{equation}
where $q_{\tau}$ is a damping parameter of the phase damping channels.
Eq.~\ref{eq:QSLT-PhaseDaming} suggest that the QSLT is independent
of acceleration parameter for the quantum systems coupled with the
phase damping channels in non-inertial frames. This acceleration-independent
behavior of QSLT is in consistent with the result obtained for the
phased-damped system of the fermionic field in non-inertial frame
\citep{Xu2020} 

\section{Conclusions\label{sec:Conclusions}}

We have investigated the speed of evolution of the amplitude damped
quantum system under Unruh decoherence from the perspective of QSLT.
For the scalar field, we have observed a speed-up of quantum evolution
of the system due to Unruh decoherence for the damping parameter $p_{\tau}\lesssim p_{\tau_{c0}}$.
The QSLT in this region turned out to be a decreasing function of
the acceleration parameter for a given damping. Moreover, on the interval
$p_{\tau}\in(p_{\tau_{c1}},p_{\tau_{c0}})$, the relativistic effect
first speeded-up, and then decelerated the quantum evolution process
of the system. On the other hand, we have noticed a deceleration quantum
evolution process for $p_{\tau_{c1}}\lesssim p_{\tau}$ in the relativistic
frame. It is important to mention that the QSLT reduced/raised to
a fixed value in the large acceleration limit. Furthermore, we have
studied the influence of relativistic effects on the QSLT for the
phase damped open quantum systems in non-inertial frames. Our results
demonstrated no relativistic effects on the speed of quantum evolution
of the system in accelerated frames.
\begin{acknowledgments}
For this work, NAK was supported by the Fundao da Cincia e Tecnologia
and COMPETE 2020 program in FEDER component (EU), through the projects
POCI-01-0145-FEDER-028887 and UID/FIS/04650/2013.
\end{acknowledgments}

\bibliographystyle{apsrev4-1}
\bibliography{rQSLT}

\begin{thebibliography}{33}%
\makeatletter
\providecommand \@ifxundefined [1]{%
 \@ifx{#1\undefined}
}%
\providecommand \@ifnum [1]{%
 \ifnum #1\expandafter \@firstoftwo
 \else \expandafter \@secondoftwo
 \fi
}%
\providecommand \@ifx [1]{%
 \ifx #1\expandafter \@firstoftwo
 \else \expandafter \@secondoftwo
 \fi
}%
\providecommand \natexlab [1]{#1}%
\providecommand \enquote  [1]{``#1''}%
\providecommand \bibnamefont  [1]{#1}%
\providecommand \bibfnamefont [1]{#1}%
\providecommand \citenamefont [1]{#1}%
\providecommand \href@noop [0]{\@secondoftwo}%
\providecommand \href [0]{\begingroup \@sanitize@url \@href}%
\providecommand \@href[1]{\@@startlink{#1}\@@href}%
\providecommand \@@href[1]{\endgroup#1\@@endlink}%
\providecommand \@sanitize@url [0]{\catcode `\\12\catcode `\$12\catcode
  `\&12\catcode `\#12\catcode `\^12\catcode `\_12\catcode `\%12\relax}%
\providecommand \@@startlink[1]{}%
\providecommand \@@endlink[0]{}%
\providecommand \url  [0]{\begingroup\@sanitize@url \@url }%
\providecommand \@url [1]{\endgroup\@href {#1}{\urlprefix }}%
\providecommand \urlprefix  [0]{URL }%
\providecommand \Eprint [0]{\href }%
\providecommand \doibase [0]{http://dx.doi.org/}%
\providecommand \selectlanguage [0]{\@gobble}%
\providecommand \bibinfo  [0]{\@secondoftwo}%
\providecommand \bibfield  [0]{\@secondoftwo}%
\providecommand \translation [1]{[#1]}%
\providecommand \BibitemOpen [0]{}%
\providecommand \bibitemStop [0]{}%
\providecommand \bibitemNoStop [0]{.\EOS\space}%
\providecommand \EOS [0]{\spacefactor3000\relax}%
\providecommand \BibitemShut  [1]{\csname bibitem#1\endcsname}%
\let\auto@bib@innerbib\@empty
\bibitem [{\citenamefont {Fuentes-Schuller}\ and\ \citenamefont
  {Mann}(2005)}]{Fuentes-Schuller2005}%
  \BibitemOpen
  \bibfield  {author} {\bibinfo {author} {\bibfnamefont {I.}~\bibnamefont
  {Fuentes-Schuller}}\ and\ \bibinfo {author} {\bibfnamefont {R.~B.}\
  \bibnamefont {Mann}},\ }\href {\doibase 10.1103/PhysRevLett.95.120404}
  {\bibfield  {journal} {\bibinfo  {journal} {Phys. Rev. Lett.}\ }\textbf
  {\bibinfo {volume} {95}},\ \bibinfo {pages} {120404} (\bibinfo {year}
  {2005})}\BibitemShut {NoStop}%
\bibitem [{\citenamefont {Alsing}\ \emph {et~al.}(2006)\citenamefont {Alsing},
  \citenamefont {Fuentes-Schuller}, \citenamefont {Mann},\ and\ \citenamefont
  {Tessier}}]{Alsing_2006}%
  \BibitemOpen
  \bibfield  {author} {\bibinfo {author} {\bibfnamefont {P.~M.}\ \bibnamefont
  {Alsing}}, \bibinfo {author} {\bibfnamefont {I.}~\bibnamefont
  {Fuentes-Schuller}}, \bibinfo {author} {\bibfnamefont {R.~B.}\ \bibnamefont
  {Mann}}, \ and\ \bibinfo {author} {\bibfnamefont {T.~E.}\ \bibnamefont
  {Tessier}},\ }\href {\doibase 10.1103/PhysRevA.74.032326} {\bibfield
  {journal} {\bibinfo  {journal} {Phys. Rev. A}\ }\textbf {\bibinfo {volume}
  {74}},\ \bibinfo {pages} {032326} (\bibinfo {year} {2006})}\BibitemShut
  {NoStop}%
\bibitem [{\citenamefont {Bruschi}\ \emph {et~al.}(2010)\citenamefont
  {Bruschi}, \citenamefont {Louko}, \citenamefont {Mart\'{\i}n-Mart\'{\i}nez},
  \citenamefont {Dragan},\ and\ \citenamefont {Fuentes}}]{Bruschi2010}%
  \BibitemOpen
  \bibfield  {author} {\bibinfo {author} {\bibfnamefont {D.~E.}\ \bibnamefont
  {Bruschi}}, \bibinfo {author} {\bibfnamefont {J.}~\bibnamefont {Louko}},
  \bibinfo {author} {\bibfnamefont {E.}~\bibnamefont
  {Mart\'{\i}n-Mart\'{\i}nez}}, \bibinfo {author} {\bibfnamefont
  {A.}~\bibnamefont {Dragan}}, \ and\ \bibinfo {author} {\bibfnamefont
  {I.}~\bibnamefont {Fuentes}},\ }\href {\doibase 10.1103/PhysRevA.82.042332}
  {\bibfield  {journal} {\bibinfo  {journal} {Phys. Rev. A}\ }\textbf {\bibinfo
  {volume} {82}},\ \bibinfo {pages} {042332} (\bibinfo {year}
  {2010})}\BibitemShut {NoStop}%
\bibitem [{\citenamefont {Wen}\ and\ \citenamefont {Wang}(2014)}]{Wen_2014}%
  \BibitemOpen
  \bibfield  {author} {\bibinfo {author} {\bibfnamefont {C.-H.}\ \bibnamefont
  {Wen}}\ and\ \bibinfo {author} {\bibfnamefont {J.-C.}\ \bibnamefont {Wang}},\
  }\href {\doibase 10.1088/0253-6102/62/3/06} {\bibfield  {journal} {\bibinfo
  {journal} {Communications in Theoretical Physics}\ }\textbf {\bibinfo
  {volume} {62}},\ \bibinfo {pages} {327} (\bibinfo {year} {2014})}\BibitemShut
  {NoStop}%
\bibitem [{\citenamefont {Khan}(2014)}]{SKSafi2014}%
  \BibitemOpen
  \bibfield  {author} {\bibinfo {author} {\bibfnamefont {S.}~\bibnamefont
  {Khan}},\ }\href {\doibase https://doi.org/10.1016/j.aop.2014.05.022}
  {\bibfield  {journal} {\bibinfo  {journal} {Annals of Physics}\ }\textbf
  {\bibinfo {volume} {348}},\ \bibinfo {pages} {270 } (\bibinfo {year}
  {2014})}\BibitemShut {NoStop}%
\bibitem [{\citenamefont {Richter}\ and\ \citenamefont
  {Omar}(2015)}]{Yasser2015}%
  \BibitemOpen
  \bibfield  {author} {\bibinfo {author} {\bibfnamefont {B.}~\bibnamefont
  {Richter}}\ and\ \bibinfo {author} {\bibfnamefont {Y.}~\bibnamefont {Omar}},\
  }\href {\doibase 10.1103/PhysRevA.92.022334} {\bibfield  {journal} {\bibinfo
  {journal} {Phys. Rev. A}\ }\textbf {\bibinfo {volume} {92}},\ \bibinfo
  {pages} {022334} (\bibinfo {year} {2015})}\BibitemShut {NoStop}%
\bibitem [{\citenamefont {Qiang}\ \emph {et~al.}(2018)\citenamefont {Qiang},
  \citenamefont {Sun}, \citenamefont {Dong},\ and\ \citenamefont
  {Dong}}]{Qiang2018}%
  \BibitemOpen
  \bibfield  {author} {\bibinfo {author} {\bibfnamefont {W.-C.}\ \bibnamefont
  {Qiang}}, \bibinfo {author} {\bibfnamefont {G.-H.}\ \bibnamefont {Sun}},
  \bibinfo {author} {\bibfnamefont {Q.}~\bibnamefont {Dong}}, \ and\ \bibinfo
  {author} {\bibfnamefont {S.-H.}\ \bibnamefont {Dong}},\ }\href {\doibase
  10.1103/PhysRevA.98.022320} {\bibfield  {journal} {\bibinfo  {journal} {Phys.
  Rev. A}\ }\textbf {\bibinfo {volume} {98}},\ \bibinfo {pages} {022320}
  (\bibinfo {year} {2018})}\BibitemShut {NoStop}%
\bibitem [{\citenamefont {Dong}\ \emph {et~al.}(2018)\citenamefont {Dong},
  \citenamefont {Torres-Arenas}, \citenamefont {Sun}, \citenamefont {Qiang},\
  and\ \citenamefont {Dong}}]{Dong2018}%
  \BibitemOpen
  \bibfield  {author} {\bibinfo {author} {\bibfnamefont {Q.}~\bibnamefont
  {Dong}}, \bibinfo {author} {\bibfnamefont {A.~J.}\ \bibnamefont
  {Torres-Arenas}}, \bibinfo {author} {\bibfnamefont {G.-H.}\ \bibnamefont
  {Sun}}, \bibinfo {author} {\bibfnamefont {W.-C.}\ \bibnamefont {Qiang}}, \
  and\ \bibinfo {author} {\bibfnamefont {S.-H.}\ \bibnamefont {Dong}},\ }\href
  {\doibase 10.1007/s11467-018-0876-x} {\bibfield  {journal} {\bibinfo
  {journal} {Frontiers of Physics}\ }\textbf {\bibinfo {volume} {14}},\
  \bibinfo {pages} {21603} (\bibinfo {year} {2018})}\BibitemShut {NoStop}%
\bibitem [{\citenamefont {Dong}\ \emph
  {et~al.}(2019{\natexlab{a}})\citenamefont {Dong}, \citenamefont
  {Torres-Arenas}, \citenamefont {Sun},\ and\ \citenamefont {Dong}}]{Dong2019}%
  \BibitemOpen
  \bibfield  {author} {\bibinfo {author} {\bibfnamefont {Q.}~\bibnamefont
  {Dong}}, \bibinfo {author} {\bibfnamefont {A.~J.}\ \bibnamefont
  {Torres-Arenas}}, \bibinfo {author} {\bibfnamefont {G.-H.}\ \bibnamefont
  {Sun}}, \ and\ \bibinfo {author} {\bibfnamefont {S.-H.}\ \bibnamefont
  {Dong}},\ }\href {https://doi.org/10.1007/s11467-019-0940-1} {\bibfield
  {journal} {\bibinfo  {journal} {Frontiers of Physics}\ }\textbf {\bibinfo
  {volume} {15}},\ \bibinfo {pages} {11602} (\bibinfo {year}
  {2019}{\natexlab{a}})}\BibitemShut {NoStop}%
\bibitem [{\citenamefont {Qiang}\ \emph {et~al.}(2019)\citenamefont {Qiang},
  \citenamefont {Dong}, \citenamefont {Mercado~Sanchez}, \citenamefont {Sun},\
  and\ \citenamefont {Dong}}]{Qiang2019}%
  \BibitemOpen
  \bibfield  {author} {\bibinfo {author} {\bibfnamefont {W.-C.}\ \bibnamefont
  {Qiang}}, \bibinfo {author} {\bibfnamefont {Q.}~\bibnamefont {Dong}},
  \bibinfo {author} {\bibfnamefont {M.~A.}\ \bibnamefont {Mercado~Sanchez}},
  \bibinfo {author} {\bibfnamefont {G.-H.}\ \bibnamefont {Sun}}, \ and\
  \bibinfo {author} {\bibfnamefont {S.-H.}\ \bibnamefont {Dong}},\ }\href
  {https://doi.org/10.1007/s11128-019-2421-4} {\bibfield  {journal} {\bibinfo
  {journal} {Quantum Information Processing}\ }\textbf {\bibinfo {volume}
  {18}},\ \bibinfo {pages} {314} (\bibinfo {year} {2019})}\BibitemShut
  {NoStop}%
\bibitem [{\citenamefont {Dong}\ \emph
  {et~al.}(2019{\natexlab{b}})\citenamefont {Dong}, \citenamefont {Sanchez},
  \citenamefont {Sun}, \citenamefont {Toutounji},\ and\ \citenamefont
  {Dong}}]{Dong_2019}%
  \BibitemOpen
  \bibfield  {author} {\bibinfo {author} {\bibfnamefont {Q.}~\bibnamefont
  {Dong}}, \bibinfo {author} {\bibfnamefont {M.~A.~M.}\ \bibnamefont
  {Sanchez}}, \bibinfo {author} {\bibfnamefont {G.-H.}\ \bibnamefont {Sun}},
  \bibinfo {author} {\bibfnamefont {M.}~\bibnamefont {Toutounji}}, \ and\
  \bibinfo {author} {\bibfnamefont {S.-H.}\ \bibnamefont {Dong}},\ }\href
  {\doibase 10.1088/0256-307x/36/10/100301} {\bibfield  {journal} {\bibinfo
  {journal} {Chinese Physics Letters}\ }\textbf {\bibinfo {volume} {36}},\
  \bibinfo {pages} {100301} (\bibinfo {year} {2019}{\natexlab{b}})}\BibitemShut
  {NoStop}%
\bibitem [{\citenamefont {Khan}\ \emph {et~al.}(2020)\citenamefont {Khan},
  \citenamefont {Amin},\ and\ \citenamefont {Jan}}]{Niaz2020}%
  \BibitemOpen
  \bibfield  {author} {\bibinfo {author} {\bibfnamefont {N.~A.}\ \bibnamefont
  {Khan}}, \bibinfo {author} {\bibfnamefont {S.~T.}\ \bibnamefont {Amin}}, \
  and\ \bibinfo {author} {\bibfnamefont {M.}~\bibnamefont {Jan}},\ }\href
  {https://doi.org/10.1088/1572-9494/abbccf} {\bibfield  {journal} {\bibinfo
  {journal} {Commun. Theor. Phys.}\ }\textbf {\bibinfo {volume} {72}},\
  \bibinfo {pages} {125103} (\bibinfo {year} {2020})}\BibitemShut {NoStop}%
\bibitem [{\citenamefont {Shamsi}\ \emph {et~al.}(2020)\citenamefont {Shamsi},
  \citenamefont {Noreen},\ and\ \citenamefont {Mushtaq}}]{Shamsi2020}%
  \BibitemOpen
  \bibfield  {author} {\bibinfo {author} {\bibfnamefont {Z.~H.}\ \bibnamefont
  {Shamsi}}, \bibinfo {author} {\bibfnamefont {A.}~\bibnamefont {Noreen}}, \
  and\ \bibinfo {author} {\bibfnamefont {A.}~\bibnamefont {Mushtaq}},\ }\href
  {\doibase https://doi.org/10.1016/j.rinp.2020.103302} {\bibfield  {journal}
  {\bibinfo  {journal} {Results in Physics}\ }\textbf {\bibinfo {volume}
  {19}},\ \bibinfo {pages} {103302} (\bibinfo {year} {2020})}\BibitemShut
  {NoStop}%
\bibitem [{\citenamefont {Sanchez}\ \emph {et~al.}(2020)\citenamefont
  {Sanchez}, \citenamefont {Sun},\ and\ \citenamefont {Dong}}]{Sanchez_2020}%
  \BibitemOpen
  \bibfield  {author} {\bibinfo {author} {\bibfnamefont {M.~A.~M.}\
  \bibnamefont {Sanchez}}, \bibinfo {author} {\bibfnamefont {G.-H.}\
  \bibnamefont {Sun}}, \ and\ \bibinfo {author} {\bibfnamefont {S.-H.}\
  \bibnamefont {Dong}},\ }\href {\doibase 10.1088/1402-4896/abbf72} {\bibfield
  {journal} {\bibinfo  {journal} {Physica Scripta}\ }\textbf {\bibinfo {volume}
  {95}},\ \bibinfo {pages} {115102} (\bibinfo {year} {2020})}\BibitemShut
  {NoStop}%
\bibitem [{\citenamefont {Emparan}(2006)}]{Emparan2006}%
  \BibitemOpen
  \bibfield  {author} {\bibinfo {author} {\bibfnamefont {R.}~\bibnamefont
  {Emparan}},\ }\href {\doibase 10.1088/1126-6708/2006/06/012} {\bibfield
  {journal} {\bibinfo  {journal} {Journal of High Energy Physics}\ }\textbf
  {\bibinfo {volume} {2006}},\ \bibinfo {pages} {012} (\bibinfo {year}
  {2006})}\BibitemShut {NoStop}%
\bibitem [{\citenamefont {Mart\'{\i}n-Mart\'{\i}nez}\ \emph
  {et~al.}(2010)\citenamefont {Mart\'{\i}n-Mart\'{\i}nez}, \citenamefont
  {Garay},\ and\ \citenamefont {Le\'on}}]{Martin2010}%
  \BibitemOpen
  \bibfield  {author} {\bibinfo {author} {\bibfnamefont {E.}~\bibnamefont
  {Mart\'{\i}n-Mart\'{\i}nez}}, \bibinfo {author} {\bibfnamefont {L.~J.}\
  \bibnamefont {Garay}}, \ and\ \bibinfo {author} {\bibfnamefont
  {J.}~\bibnamefont {Le\'on}},\ }\href {\doibase 10.1103/PhysRevD.82.064028}
  {\bibfield  {journal} {\bibinfo  {journal} {Phys. Rev. D}\ }\textbf {\bibinfo
  {volume} {82}},\ \bibinfo {pages} {064028} (\bibinfo {year}
  {2010})}\BibitemShut {NoStop}%
\bibitem [{\citenamefont {Fuentes}\ \emph {et~al.}(2010)\citenamefont
  {Fuentes}, \citenamefont {Mann}, \citenamefont {Mart\'{\i}n-Mart\'{\i}nez},\
  and\ \citenamefont {Moradi}}]{Fuentes2010}%
  \BibitemOpen
  \bibfield  {author} {\bibinfo {author} {\bibfnamefont {I.}~\bibnamefont
  {Fuentes}}, \bibinfo {author} {\bibfnamefont {R.~B.}\ \bibnamefont {Mann}},
  \bibinfo {author} {\bibfnamefont {E.}~\bibnamefont
  {Mart\'{\i}n-Mart\'{\i}nez}}, \ and\ \bibinfo {author} {\bibfnamefont
  {S.}~\bibnamefont {Moradi}},\ }\href {\doibase 10.1103/PhysRevD.82.045030}
  {\bibfield  {journal} {\bibinfo  {journal} {Phys. Rev. D}\ }\textbf {\bibinfo
  {volume} {82}},\ \bibinfo {pages} {045030} (\bibinfo {year}
  {2010})}\BibitemShut {NoStop}%
\bibitem [{\citenamefont {Ahmadi}\ \emph {et~al.}(2014)\citenamefont {Ahmadi},
  \citenamefont {Bruschi}, \citenamefont {Sabín}, \citenamefont {Adesso},\
  and\ \citenamefont {Fuentes}}]{Ahmadi2014}%
  \BibitemOpen
  \bibfield  {author} {\bibinfo {author} {\bibfnamefont {M.}~\bibnamefont
  {Ahmadi}}, \bibinfo {author} {\bibfnamefont {D.~E.}\ \bibnamefont {Bruschi}},
  \bibinfo {author} {\bibfnamefont {C.}~\bibnamefont {Sabín}}, \bibinfo
  {author} {\bibfnamefont {G.}~\bibnamefont {Adesso}}, \ and\ \bibinfo {author}
  {\bibfnamefont {I.}~\bibnamefont {Fuentes}},\ }\href
  {https://doi.org/10.1038/srep04996} {\bibfield  {journal} {\bibinfo
  {journal} {Scientific Reports}\ }\textbf {\bibinfo {volume} {4}},\ \bibinfo
  {pages} {4996} (\bibinfo {year} {2014})}\BibitemShut {NoStop}%
\bibitem [{\citenamefont {Tian}\ \emph {et~al.}(2015)\citenamefont {Tian},
  \citenamefont {Wang}, \citenamefont {Fan},\ and\ \citenamefont
  {Jing}}]{Tian2015}%
  \BibitemOpen
  \bibfield  {author} {\bibinfo {author} {\bibfnamefont {Z.}~\bibnamefont
  {Tian}}, \bibinfo {author} {\bibfnamefont {J.}~\bibnamefont {Wang}}, \bibinfo
  {author} {\bibfnamefont {H.}~\bibnamefont {Fan}}, \ and\ \bibinfo {author}
  {\bibfnamefont {J.}~\bibnamefont {Jing}},\ }\href
  {https://doi.org/10.1038/srep07946} {\bibfield  {journal} {\bibinfo
  {journal} {Scientific Reports}\ }\textbf {\bibinfo {volume} {5}},\ \bibinfo
  {pages} {7946} (\bibinfo {year} {2015})}\BibitemShut {NoStop}%
\bibitem [{\citenamefont {Alsing}\ and\ \citenamefont
  {Milburn}(2003)}]{Alsing2003}%
  \BibitemOpen
  \bibfield  {author} {\bibinfo {author} {\bibfnamefont {P.~M.}\ \bibnamefont
  {Alsing}}\ and\ \bibinfo {author} {\bibfnamefont {G.~J.}\ \bibnamefont
  {Milburn}},\ }\href {\doibase 10.1103/PhysRevLett.91.180404} {\bibfield
  {journal} {\bibinfo  {journal} {Phys. Rev. Lett.}\ }\textbf {\bibinfo
  {volume} {91}},\ \bibinfo {pages} {180404} (\bibinfo {year}
  {2003})}\BibitemShut {NoStop}%
\bibitem [{\citenamefont {Fink}\ \emph {et~al.}(2017)\citenamefont {Fink},
  \citenamefont {Rodriguez-Aramendia}, \citenamefont {Handsteiner},
  \citenamefont {Ziarkash}, \citenamefont {Steinlechner}, \citenamefont
  {Scheidl}, \citenamefont {Fuentes}, \citenamefont {Pienaar}, \citenamefont
  {Ralph},\ and\ \citenamefont {Ursin}}]{Fink2017}%
  \BibitemOpen
  \bibfield  {author} {\bibinfo {author} {\bibfnamefont {M.}~\bibnamefont
  {Fink}}, \bibinfo {author} {\bibfnamefont {A.}~\bibnamefont
  {Rodriguez-Aramendia}}, \bibinfo {author} {\bibfnamefont {J.}~\bibnamefont
  {Handsteiner}}, \bibinfo {author} {\bibfnamefont {A.}~\bibnamefont
  {Ziarkash}}, \bibinfo {author} {\bibfnamefont {F.}~\bibnamefont
  {Steinlechner}}, \bibinfo {author} {\bibfnamefont {T.}~\bibnamefont
  {Scheidl}}, \bibinfo {author} {\bibfnamefont {I.}~\bibnamefont {Fuentes}},
  \bibinfo {author} {\bibfnamefont {J.}~\bibnamefont {Pienaar}}, \bibinfo
  {author} {\bibfnamefont {T.~C.}\ \bibnamefont {Ralph}}, \ and\ \bibinfo
  {author} {\bibfnamefont {R.}~\bibnamefont {Ursin}},\ }\href
  {https://doi.org/10.1038/ncomms15304} {\bibfield  {journal} {\bibinfo
  {journal} {Nature Communications}\ }\textbf {\bibinfo {volume} {8}},\
  \bibinfo {pages} {15304} (\bibinfo {year} {2017})}\BibitemShut {NoStop}%
\bibitem [{\citenamefont {Nagele}\ \emph {et~al.}(2020)\citenamefont {Nagele},
  \citenamefont {Ilo-Okeke}, \citenamefont {Rohde}, \citenamefont {Dowling},\
  and\ \citenamefont {Byrnes}}]{Nagele2020}%
  \BibitemOpen
  \bibfield  {author} {\bibinfo {author} {\bibfnamefont {C.}~\bibnamefont
  {Nagele}}, \bibinfo {author} {\bibfnamefont {E.~O.}\ \bibnamefont
  {Ilo-Okeke}}, \bibinfo {author} {\bibfnamefont {P.~P.}\ \bibnamefont
  {Rohde}}, \bibinfo {author} {\bibfnamefont {J.~P.}\ \bibnamefont {Dowling}},
  \ and\ \bibinfo {author} {\bibfnamefont {T.}~\bibnamefont {Byrnes}},\ }\href
  {\doibase https://doi.org/10.1016/j.physleta.2020.126301} {\bibfield
  {journal} {\bibinfo  {journal} {Physics Letters A}\ }\textbf {\bibinfo
  {volume} {384}},\ \bibinfo {pages} {126301} (\bibinfo {year}
  {2020})}\BibitemShut {NoStop}%
\bibitem [{\citenamefont {Zhang}\ \emph {et~al.}(2014)\citenamefont {Zhang},
  \citenamefont {Han}, \citenamefont {Xia}, \citenamefont {Cao},\ and\
  \citenamefont {Fan}}]{Zhang2014}%
  \BibitemOpen
  \bibfield  {author} {\bibinfo {author} {\bibfnamefont {Y.-J.}\ \bibnamefont
  {Zhang}}, \bibinfo {author} {\bibfnamefont {W.}~\bibnamefont {Han}}, \bibinfo
  {author} {\bibfnamefont {Y.-J.}\ \bibnamefont {Xia}}, \bibinfo {author}
  {\bibfnamefont {J.-P.}\ \bibnamefont {Cao}}, \ and\ \bibinfo {author}
  {\bibfnamefont {H.}~\bibnamefont {Fan}},\ }\href
  {https://doi.org/10.1038/srep04890} {\bibfield  {journal} {\bibinfo
  {journal} {Scientific Reports}\ }\textbf {\bibinfo {volume} {4}},\ \bibinfo
  {pages} {4890} (\bibinfo {year} {2014})}\BibitemShut {NoStop}%
\bibitem [{\citenamefont {Campaioli}\ \emph {et~al.}(2019)\citenamefont
  {Campaioli}, \citenamefont {Pollock},\ and\ \citenamefont
  {Modi}}]{Campaioli2019}%
  \BibitemOpen
  \bibfield  {author} {\bibinfo {author} {\bibfnamefont {F.}~\bibnamefont
  {Campaioli}}, \bibinfo {author} {\bibfnamefont {F.}~\bibnamefont {Pollock}},
  \ and\ \bibinfo {author} {\bibfnamefont {K.}~\bibnamefont {Modi}},\
  }\href@noop {} {\bibfield  {journal} {\bibinfo  {journal} {Quantum}\ }\textbf
  {\bibinfo {volume} {3}} (\bibinfo {year} {2019})}\BibitemShut {NoStop}%
\bibitem [{\citenamefont {Musadiq}\ \emph {et~al.}(2019)\citenamefont
  {Musadiq}, \citenamefont {Khan}, \citenamefont {Javed},\ and\ \citenamefont
  {Shamirzaie}}]{Musadiq2019}%
  \BibitemOpen
  \bibfield  {author} {\bibinfo {author} {\bibfnamefont {M.}~\bibnamefont
  {Musadiq}}, \bibinfo {author} {\bibfnamefont {S.}~\bibnamefont {Khan}},
  \bibinfo {author} {\bibfnamefont {M.}~\bibnamefont {Javed}}, \ and\ \bibinfo
  {author} {\bibfnamefont {M.}~\bibnamefont {Shamirzaie}},\ }\href {\doibase
  10.1142/S0219749919500540} {\bibfield  {journal} {\bibinfo  {journal}
  {International Journal of Quantum Information}\ }\textbf {\bibinfo {volume}
  {17}},\ \bibinfo {pages} {1950054} (\bibinfo {year} {2019})}\BibitemShut
  {NoStop}%
\bibitem [{\citenamefont {Musadiq}\ and\ \citenamefont
  {Khan}(2020)}]{Musadiq2020}%
  \BibitemOpen
  \bibfield  {author} {\bibinfo {author} {\bibfnamefont {M.}~\bibnamefont
  {Musadiq}}\ and\ \bibinfo {author} {\bibfnamefont {S.}~\bibnamefont {Khan}},\
  }\href {\doibase 10.1088/1751-8121/abc21e} {\bibfield  {journal} {\bibinfo
  {journal} {Journal of Physics A: Mathematical and Theoretical}\ }\textbf
  {\bibinfo {volume} {53}},\ \bibinfo {pages} {505302} (\bibinfo {year}
  {2020})}\BibitemShut {NoStop}%
\bibitem [{\citenamefont {Khan}\ and\ \citenamefont {Khan}(2015)}]{Niaz2015}%
  \BibitemOpen
  \bibfield  {author} {\bibinfo {author} {\bibfnamefont {S.}~\bibnamefont
  {Khan}}\ and\ \bibinfo {author} {\bibfnamefont {N.~A.}\ \bibnamefont
  {Khan}},\ }\href {https://doi.org/10.1140/epjp/i2015-15216-0} {\bibfield
  {journal} {\bibinfo  {journal} {The European Physical Journal Plus}\ }\textbf
  {\bibinfo {volume} {130}},\ \bibinfo {pages} {216} (\bibinfo {year}
  {2015})}\BibitemShut {NoStop}%
\bibitem [{\citenamefont {Haseli}(2019)}]{Haseli2019}%
  \BibitemOpen
  \bibfield  {author} {\bibinfo {author} {\bibfnamefont {S.}~\bibnamefont
  {Haseli}},\ }\href {https://doi.org/10.1140/epjc/s10052-019-7129-1}
  {\bibfield  {journal} {\bibinfo  {journal} {The European Physical Journal C}\
  }\textbf {\bibinfo {volume} {79}},\ \bibinfo {pages} {616} (\bibinfo {year}
  {2019})}\BibitemShut {NoStop}%
\bibitem [{\citenamefont {Xu}\ \emph {et~al.}(2020)\citenamefont {Xu},
  \citenamefont {Zhu}, \citenamefont {Zhang}, \citenamefont {Wang},\ and\
  \citenamefont {Liu}}]{Xu2020}%
  \BibitemOpen
  \bibfield  {author} {\bibinfo {author} {\bibfnamefont {K.}~\bibnamefont
  {Xu}}, \bibinfo {author} {\bibfnamefont {H.}~\bibnamefont {Zhu}}, \bibinfo
  {author} {\bibfnamefont {G.}~\bibnamefont {Zhang}}, \bibinfo {author}
  {\bibfnamefont {J.}~\bibnamefont {Wang}}, \ and\ \bibinfo {author}
  {\bibfnamefont {W.}~\bibnamefont {Liu}},\ }\href@noop {} {\bibfield
  {journal} {\bibinfo  {journal} {The European Physical Journal C}\ }\textbf
  {\bibinfo {volume} {80}},\ \bibinfo {pages} {1} (\bibinfo {year}
  {2020})}\BibitemShut {NoStop}%
\bibitem [{\citenamefont {Takagi}(1986)}]{Takagi1986}%
  \BibitemOpen
  \bibfield  {author} {\bibinfo {author} {\bibfnamefont {S.}~\bibnamefont
  {Takagi}},\ }\href {\doibase 10.1143/PTP.88.1} {\bibfield  {journal}
  {\bibinfo  {journal} {Progress of Theoretical Physics Supplement}\ }\textbf
  {\bibinfo {volume} {88}},\ \bibinfo {pages} {1} (\bibinfo {year}
  {1986})}\BibitemShut {NoStop}%
\bibitem [{\citenamefont {Sun}\ and\ \citenamefont {Zheng}(2019)}]{Sun2019}%
  \BibitemOpen
  \bibfield  {author} {\bibinfo {author} {\bibfnamefont {S.}~\bibnamefont
  {Sun}}\ and\ \bibinfo {author} {\bibfnamefont {Y.}~\bibnamefont {Zheng}},\
  }\href {\doibase 10.1103/PhysRevLett.123.180403} {\bibfield  {journal}
  {\bibinfo  {journal} {Phys. Rev. Lett.}\ }\textbf {\bibinfo {volume} {123}},\
  \bibinfo {pages} {180403} (\bibinfo {year} {2019})}\BibitemShut {NoStop}%
\bibitem [{\citenamefont {Nielsen}\ and\ \citenamefont
  {Chuang}(2000)}]{Nielsen2000}%
  \BibitemOpen
  \bibfield  {author} {\bibinfo {author} {\bibfnamefont {M.~A.}\ \bibnamefont
  {Nielsen}}\ and\ \bibinfo {author} {\bibfnamefont {I.~L.}\ \bibnamefont
  {Chuang}},\ }\href@noop {} {\emph {\bibinfo {title} {Quantum computation and
  quantum information}}}\ (\bibinfo  {publisher} {Cambridge Univ. Press},\
  \bibinfo {year} {2000})\BibitemShut {NoStop}%
\bibitem [{Note1()}]{Note1}%
  \BibitemOpen
  \bibinfo {note} {The integral has the simplified form \begin {align*} -\DOTSI
  \intop \ilimits@ _{1}^{p_{\tau }}dp_{t}\protect \sqrt {\protect \frac
  {4p_{t}+a^{2}(r)}{p_{t}}} & =\protect \sqrt {4+a^{2}(r)}-\protect \sqrt
  {p_{\tau }\left (4p_{\tau }+a^{2}(r)\right )}\\ & -\protect \frac
  {1}{2}a^{2}(r)(\protect \qopname \relax o{sinh}^{-1}\protect \frac {2\protect
  \sqrt {p_{\tau }}}{a(r)}-\protect \qopname \relax o{sinh}^{-1}\protect \frac
  {2}{a(r)}). \end {align*}}\BibitemShut {NoStop}%
\end{thebibliography}%

\end{document}